\documentclass[10pt,conference]{IEEEtran}
\IEEEoverridecommandlockouts
\usepackage{cite}
\usepackage{amsmath,amssymb,amsfonts}
\usepackage{algorithmic}
\usepackage{graphicx}
\usepackage{textcomp}
\usepackage{xcolor}
\def\BibTeX{{\rm B\kern-.05em{\sc i\kern-.025em b}\kern-.08em
    T\kern-.1667em\lower.7ex\hbox{E}\kern-.125emX}}

\usepackage{graphicx} 
\usepackage[caption=false,font=normalsize,labelfont=sf,textfont=sf]{subfig}

\usepackage{xcolor,soul,framed} 
\usepackage{xspace}
\usepackage{algorithm}  
\usepackage{algorithmic}
\usepackage{balance}
\usepackage{amsmath}
\usepackage{titlesec}
\usepackage[toc]{glossaries}

\newcommand{\ie}{{\em i.e.,}\ }
\newcommand{\eg}{{\em e.g.,}\ }

\newcommand{\etal}{{\em et al.}\ }

\renewcommand{\algorithmiccomment}[1]{\bgroup\hfill//~#1\egroup}

\makeatletter
\def\algbackskip{\hskip-\ALG@thistlm}
\makeatother

\usepackage{etoolbox}
\usepackage{xstring}
\DeclareListParser{\doslashlist}{/}
\newcounter{ndnNameComponentCounter}%
\newcommand{\name}[1]{{%
  \setcounter{ndnNameComponentCounter}{0}%
  \renewcommand{\do}[1]{{%
    \ifnumgreater{\value{ndnNameComponentCounter}}{0}{\allowbreak/}{}%
    \ifnumodd{\value{ndnNameComponentCounter}}{}{}%
    ##1}%
    \stepcounter{ndnNameComponentCounter}}%
``{\fontfamily{cmtt}\small\selectfont\IfBeginWith{#1}{/}{/}{}\doslashlist{#1}}''%
}}

\newcommand{\sol}{{\it SAMBA}\xspace}
\newcommand{\SL}{{Self Learning}\xspace}

\newcommand{\npm}{{IPA}\xspace}
\newcommand{\pfs}{{AF}\xspace}
\newcommand{\NPM}{{Implicit Prefix Aggregation}\xspace}

\newcommand{\PFS}{{Approximate Forwarding}\xspace}

\newcommand\blfootnote[1]{%
  \begingroup
  \renewcommand\thefootnote{}\footnote{#1}%
  \addtocounter{footnote}{-1}%
  \endgroup
}

\title{\sol: Scalable Approximate Forwarding For NDN Implicit FIB Aggregation}

\usepackage{caption}

\begin{document}

\author{
\IEEEauthorblockN{Amir Esmaeili}
\IEEEauthorblockA{\textit{Department of Computer Science} \\
\textit{Binghamton University}\\
Binghamton, USA \\
aesmaeili@binghamton.edu}
\and
\IEEEauthorblockN{Abderrahmen Mtibaa}
\IEEEauthorblockA{\textit{Department of Computer Science} \\
\textit{University of Missouri St. Louis}\\
St. Louis, USA \\
amtibaa@umsl.edu}
}

\maketitle

\begin{abstract}
\blfootnote{\color{red}This paper is accepted and proceeding on 2024 IEEE 13th International Conference on Cloud Networking (CloudNet).}


The Internet landscape has witnessed a significant shift toward Information Centric Networking (ICN) due to the exponential growth of data-driven applications. Similar to routing tables in TCP/IP architectures, ICN uses Forward Information Base (FIB) tables. However, FIB tables can grow exponentially due to their URL-like naming scheme, introducing major delays in the prefix lookup process. Existing explicit FIB aggregation solutions are very complex to run, and ICN on-demand routing schemes, which use a discovery mechanism to help reduce the number of FIB records and thus have shorter lookup times, rely on flooding-based mechanisms and building routes for all requests, introducing additional scalability challenges. In this paper, we propose \sol, an \PFS-based \SL, that uses the nearest FIB trie record to the given prefix for reducing the number of discoveries thus keeping the FIB table small. By choosing the nearest prefix to a given name prefix, \sol uses \NPM (\npm) which implicitly aggregates the FIB records and reduces the number of Self Learning discoveries required. Coupled with the approximate forwarding, \sol can achieve efficient and scalable forwarding.

 
We demonstrate that \sol can help reduce lookup times by up to 45\%. \sol also implements multipath discovery and consumer-controlled flooding mechanisms, which help minimize networking overhead. Our simulation results show that \sol reduces the FIB table size twenty fold compared to traditional \SL schemes.

\end{abstract}


\section{Introduction}

In the evolving landscape of network architectures, Information-Centric Networking (ICN) paradigm and its most prominent architecture Named Data Networking (NDN) architecture represents a paradigm shift that prioritizes data retrieval based on content rather than traditional host-based addressing~\cite{zhang2014named}. Central to the efficiency and effectiveness of ICN/NDN is the routing mechanism, which underpins the entire framework by finding and storing routes at each node. NDN consumers send {\em interests} to fetch any Data provided by a producer or stored in the network (\eg CDNs). ICN/NDN uses a Forward Information Based (FIB) to store routes computed by a routing algorithm, which can be used by NDN forwarding plane to send and forward each interest message towards its producer~\cite{tariq2019forwarding}.



Scalability is a significant challenge in NDN due to the nature of the Forwarding Information Base (FIB) tables. Unlike traditional IP routing tables, which manage routes based on relatively stable, hierarchical IP addresses, FIB tables however must handle a vast and dynamic range of content names~\cite{shubbar2019efficient}. These names are often hierarchical and much longer, leading to an exponential increase in the size and complexity of the FIB entries. As a result, the memory and processing power required to manage these tables grow significantly, making them less scalable. Additionally, the frequent addition and deletion of content names exacerbate the challenge, as FIB tables must dynamically adapt to these changes in real-time. Thus, developing scalable solutions for FIB management is crucial to achieving efficient and robust performance.

Researchers have been addressing this scalability challenge, by proposing: (i) optimized data structures for faster FIB lookups~\cite{karrakchou2020fctrees,song2015scalable,hu2019composite,shubbar2019efficient}, (ii) 
compression and FIB aggregation mechanisms to reduce name prefix sizes~\cite{niu2021entropy,zhang2019note,niu2021entropy,shi2017broadcast}, (iii)  
hardware-based FIB implementation for faster lookups~\cite{karrakchou2020endn,miguel2018named}, (iv) on-demand routing schemes which discover routes as needed resulting in  smaller FIB table sizes~\cite{shi2017broadcast,liang2020enabling,meisel2010listen} when compared to proactive routing approaches~\cite{wang2012ospfn,hoque2013nlsr}.
 These solutions are not scalable due to collision (i), adding an extra layer of complexity (ii), requiring specific hardware (iii), perform an explicit offline aggregation (ii), or do not scale well under high demand.




In this paper, we propose \sol, a scalable NDN forwarding system that uses an \PFS algorithm (\pfs) to forward interests based on a best-effort fashion, and when forwarding fails, an optimized \SL discovery is initiated to find and store as many routes to the producers on the reverse path. Therefore, \sol employs an \NPM (\npm) to keep the size of the FIB table small, along with the \PFS algorithm (\pfs) which performs a Longest Prefix Match (LPM) first, and when that fails sol chooses the first available FIB record on the subtrie. 
\sol's forwarding system creates an Implicit Prefix
Aggregation (\npm), which implicitly aggregates the FIB entries, unlike state-of-the-art explicit FIB aggregation schemes which incur an extra computation overhead to aggregate or compress the FIB records~\cite{niu2021entropy,shi2017broadcast}, or predictive forwarding approaches~\cite{song2015scalable,chan2017fuzzy} that can not recover the correct path when wrong forwarding happens, \sol discovers correct path via broadcasting.


\noindent {\bf Our Contributions:} In this paper, we make the following contributions:
\begin{itemize}
    \item We empirically measure the potential gains of reducing the FIB size on every router lookup operation.
    \item We propose \sol, \PFS that uses the nearest FIB record to the given prefix to avoid unnecessary route discoveries, resulting in reducing the FIB size, thus it provides an 
    \npm. It also can recover wrong paths via broadcasting.
    
    \item \sol also implements multipath and a Stop-and-Wait features, which help build alternate routes and reduce flooding in the network respectively. 
    \item Our simulation results compare the performance of \sol to state-of-the-art \SL, and shows major improvements, including reducing FIB size by up to $20\times$, overhead by 75\%, and up to  50\% more throughput during link failures. 
\end{itemize}

The remainder of this paper is organized as follows. In 
 Section~\ref{relatedwork}, we provide a  survey of related works schemes. Section~\ref{sec:prob} motivates the need for controlling the FIB table size. We present the detailed design of \sol in Section~\ref{secNPM}. Section~\ref{seceval} evaluates \sol's performance. Section~\ref{conclusion} concludes the paper and presents future research directions.

\section{Related Work} \label{relatedwork}

Routing and forwarding are two important modules in the ICN/NDN, while routing finds available routes for destinations and fills FIB table, forwarding selects the next hop for a given incoming interest from FIB table~\cite{mansour2020load}. A key function in forwarding that operates over the FIB table for every incoming interest is the lookup function. The performance of the lookup function depends on the number of FIB entries, making FIB scalability a major challenge.

FIB scalability in NDN has been explored, and there are five research categories in this domain: (i) optimized data structures for faster FIB lookups~\cite{karrakchou2020fctrees,song2015scalable,hu2019composite,shubbar2019efficient}, (ii) 
compression and FIB aggregation mechanisms to reduce name prefix sizes~\cite{zhang2019note,niu2021entropy}, (iii)  
hardware-based FIBs~\cite{karrakchou2020endn,miguel2018named}, (iv) on-demand routing schemes~\cite{shi2017broadcast,liang2020enabling,meisel2010listen} which discover routes as needed resulting in smaller FIB table sizes when compared to proactive routing approaches~\cite{wang2012ospfn,hoque2013nlsr,brito2021ndvr}, and (v) predictive forwarding~\cite{song2015scalable,chan2017fuzzy} that speculatively forwards an interest in the absence of a longest prefix match. 

However, while optimized data structures (i) such as hash-based solutions~\cite{shubbar2019efficient} are not scalable due collisions, hardware-based methods (ii), though fast in operation, are complex to implement and incur an additional cost. On the other hand, prefix compression, and coding solutions~\cite{zhang2019note,niu2021entropy} operate for each interest received, incurring a complex computation at the routers. 
Aggregation-based methods~\cite{karrakchou2020fctrees}, however, run periodically on the entire FIB tables, which cab complex, slow, and inefficient. We, in this paper, propose \sol that uses an implicit aggregation method which reuses ``similar'' FIB records keeping the FIB tables small in size.

On-demand routing (iv) and speculative forwarding (v) are critical areas of research that closely align with the focus of our work. On-demand schemes (iv)~\cite{wang2018secure,meisel2010listen,baccelli2014information,shi2017broadcast,liang2020enabling}, on the other hand, compute paths reactively (\ie on-demand, when needed), by broadcasting a discovery interest to establish a route at the network for the requested producer. On-demand schemes keep the FIB size ``smaller'' and ``manageable'' by storing only necessary routes. 
\SL~\cite{shi2017broadcast}, one of the most popular on-demand routing schemes uses a {\em discovery mechanism} to find new routes (when ``regular'' interest fail to find one) and store these routes on all nodes on the path. 
While \SL 
approaches~\cite{shi2017broadcast,liang2020enabling} reduce the number of FIB entries at the routers' FIB tables, these tables can still grow as the number of requests scale. Finally, forwarding based on speculation is another approach to keep the FIB smaller. Fuzzy forwarding~\cite{chan2017fuzzy} exploits semantic similarity to find corresponding records which is highly complex to construct in vector space, and speculative forwarding~\cite{song2015scalable} uses a token-relaxed Patricia trie as a data structure and forwards interests speculatively. This method, however, fails to address the challenge of finding a route when speculative forwarding fails to reach the producer. 

We propose \sol, which utilizes the features in both \SL and \PFS(\pfs) to forward interests to the closest FIB record for any given prefix, however when forwarding fails, an updated \SL discovery is initiated to discover as many routes as possible. \sol is designed to reduce the number of ``unnecessary'' discoveries to maintain the FIB tables as small as possible. 


\section{Why FIB Size Matters?}
\label{sec:prob}

To quantify the impact of the increasing number of FIB entries at the routers on lookup and insertion operation delays, we perform a set of benchmarking experiments using a desktop machine with an Intel core i5 CPU and 8GB of RAM. We implement the FIB data structure as a trie~\cite{liu2017unified}\footnote{Tries are known to perform faster lookups than hierarchical hash tables~\cite{karrakchou2020fctrees}}. We generate tries that consist of FIB entries for prefixes of lengths up to 50 characters. We implement our trie such that each node represents a single character. This makes our trie storing prefixes with heights up to 50. 

\begin{figure}[tbp]
    \centering
    \includegraphics[width=0.5\linewidth]{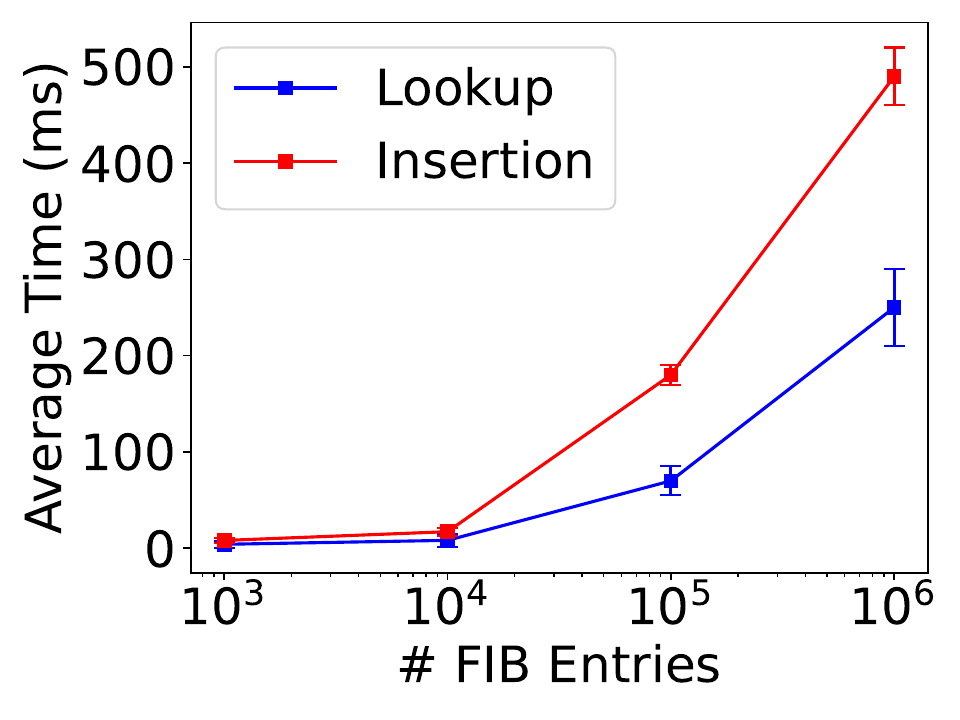}
    \caption{Scalability assessment of FIB lookup time; average of 100 FIB lookup and insertion times for as the FIB table size scales from 1k to 1M entries (x axis in logscale)}
    \label{lookuptime}
\end{figure}

 We vary the size of the trie from 1K to 1M entries and perform an average of 100 randomly generated prefix lookups and insertions. Each lookup and insertion operation is repeated 20 times and averaged. Results are shown in Figure~\ref{lookuptime}.

Figure~\ref{lookuptime} shows that as the number of FIB entries increases, the lookup and insertion time increases almost linearly. Because tries are different from binary trees, their lookup time is larger and does not scale well as the number of FIB entries grows~\cite{al1984algorithms}. 
Note that while the exact lookup and insertion time may vary depending on the implementation, the hardware, the language, etc., we are more interested in the trend showing the potential impact of a  reduction in FIB table size on the performance. For instance, the figure shows that a FIB size reduction of $10\times$, say from 1M to 100K entries can reduce the lookup time by up to 45\%. This time saved will be applied to every single interest received by the router resulted in major performance improvement across the network and a much better quality of experience for the end consumers. This paper aims to achieve smaller FIB by using the minimum number of FIB entries which is called \NPM (\npm) while maintaining the same (\ie or similar) routing/forwarding performance. 

\section{\sol} \label{secNPM}
In this section, we present \sol, a scalable approximate forwarding using enhanced \SL-based routing scheme that reduces the number of discoveries while maintaining the FIB tables sizes as small as possible. We first argue why self learning schemes are more suited to be coupled with an approximate forwarding for faster forwarding at scale.



 
\subsection{Why LPM is Not Suited for Self Learning Schemes?}
Longest prefix matching (LPM) is used by most ICN/NDN routers to determine the list of next hops for any given prefix (\ie producer). While LPM is very efficient and fast for most routing schemes, in \SL, routes are set on-demand, which means there may not be a route for all legitimate prefixes. In this case, LPM will fail to find a list of next hop faces,  triggering a discovery for a new route. {\em We argue that this process is not optimal and can introduce unnecessary route discoveries, resulting in larger FIB tables and network overhead}. In fact, often producers implement multiple services, or a server may host multiple tenant and/or multiple stakeholders services, thus a partial name prefix of the producer or the server can suffice to serve many prefixes without the need to discover all or/and store them all in the FIB~\cite{shi2017broadcast}.

\begin{figure}
     \centering
         \includegraphics[width=0.7\linewidth]{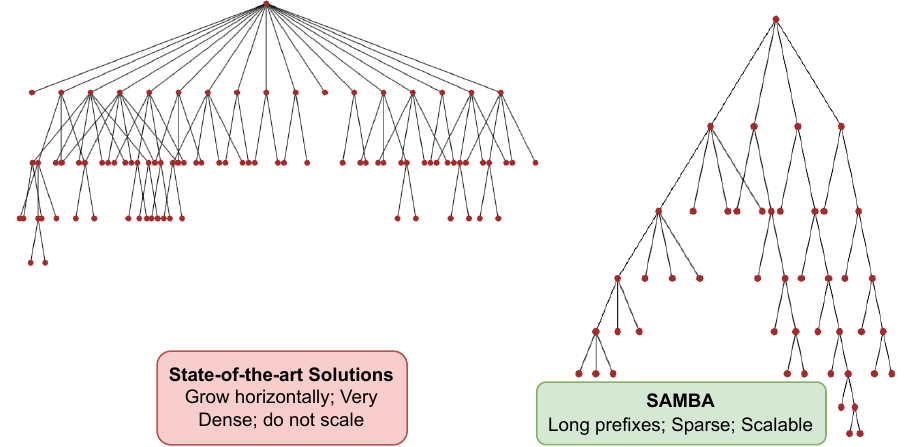}       
     \caption{Comparing the FIB properties using state-of-the-art solutions~\cite{shi2017broadcast} (left) and \sol (right). \sol inserts very few prefix entries, often longer but act as an implicit aggregate/summarization for the entire trie branch.}
     \label{fig.NPM}
 \end{figure}

For simplicity, let us assume that a given prefix $p$, of length $k$, has the following form: \name{/<domain1/domain2/.../domaink}.
$p$ is stored in the FIB as a branch in the trie of height $k$, where the leaf, \name{domaink} includes a pointer to the outgoing interface(s) for any interests matching $p$. Self Learning schemes, similar to proactive routing schemes, sets records in the FIB for all prefixes, up to \name{domaink}, which creates a very dense FIB trie structure as shown in Fig.~\ref{fig.NPM} (subfigure in the left), while \sol implicitly aggregates FIB records and re-utilizes existing similar records for new incoming interests which creates a sparse and easy to search trie structure as per the right subfigure in Fig.~\ref{fig.NPM}. For instance, if a node receives a new interest \name{/<domain1/domain2'/...}, it will first try the route set by the record \name{/<domain1/domain2/.../domaink}, however, if that routes fails then and only then the node initiates a new discovery and stores a new route. This simple \npm mechanism reduces the trie size considerably using an implicit FIB aggregation mechanism which is lightweight compared to an explicit offline aggregation which operates on the entire FIB and can be very costly~\cite{shi2017broadcast}.

Besides this simple idea, throughout the paper, we will address, analyze, and discuss the following research questions:
\begin{enumerate} 
    \item What is the complexity introduced by the new \pfs algorithm compared to the well studied and optimized longest prefix matching algorithm~\cite{shi2017broadcast}? 
     \item What is the additional delay cost introduced by \sol when it fails to reuse FIB records? 
     \item Can \npm provide similar compression performance compared to explicit FIB aggregation? 
\end{enumerate}


%
\subsection{\PFS(\pfs) \& \NPM (\npm) } \label{NPMALG}
\begin{figure}
     \centering
         \includegraphics[width=0.60\linewidth]{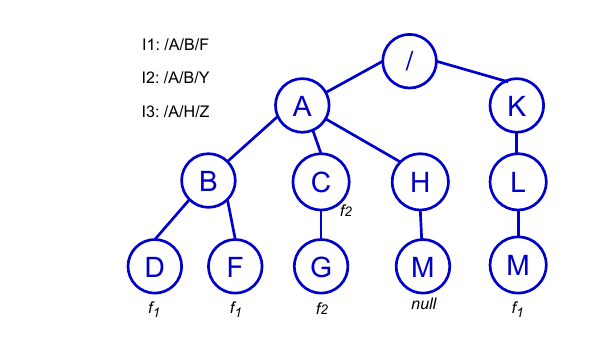}       
     \caption{Trie example: (I1) \pfs performs similar to LPM and returns $f_1$; (I2), \pfs runs a DFS on the sub-trie rooted at B and returns $f_1$ as it finds leaf node D; (I3) \pfs runs DFS on the sub-trie rooted at H which fails to find a next hop, thus \pfs returns a {\tt NOROUTE}.  }
     \label{fig.trie}
 \end{figure}

Consider a trie, a string-indexed look-up data structure, consisting of a set of nodes. All the children of a node have a common prefix of the string associated with that parent node. The trie is rooted \name{/} (refer to the trie example consisting of 6 leaf nodes in figure~\ref{fig.trie}). To determine the next hops for any given prefix $p$, The \pfs algorithm works as follows: (i) First it operates similar to the longest prefix matching (LPM) algorithm by searching from the root of the trie \name{/} until it reaches a leaf node, \ie a node in the trie which stores next hop faces for the given prefix, or a normal trie node, which we will refer to as stopping node (Sn). (ii) if a leaf node is reached then \pfs returned exactly the same next hop than LPM, however \pfs operates differently when Sn is reached--\ie LPM fails to find a next hop face. \pfs performs a simple lightweight depth-first search (DFS), searching for a next hop in the sub-trie rooted at \name{Sn}. Note that the DFS does not search the strings but simply checks if any next hops exists in the sub-trie. (iii) If DFS successfully found a list of next hop faces, say $f$, then \pfs returns $f$ as a forwarding face unless $f$ is a {\tt local} face, then \pfs returns a {\tt NOROUTE NACK} (Alg.~\ref{NPM}). This message triggers a new discovery message in consuemr side to discover the correct route. 

Thus, the interest will follow an approximate path towards a given producer; if the correct producer is reached then data is sent back to consumer(s), otherwise a given producer receiving the interest while unable to satisfy it (\ie when DFS returns $f$ which is a {\tt local} face, the interest is about to be sent to the incorrect producer application), sends a {\tt NOROUTE NACK} to the consumer. This \pfs's NACK will trigger a {\em discovery interest} by the consumer and a new prefix $p$ for the producer will be added to all nodes in the path, similar to the \SL mechanism.

To illustrate how \pfs works, we depict in Figure~\ref{fig.trie} a sample trie at a given node, $u$, consisting of six leaf nodes (\ie nodes with a next hop face, $f_i \neq null$). If $u$ receives the interest \name{/A/B/F} (I1), \pfs performs a longest prefix match (LPM) and finds a next hop face, $f1$, at node F. However, if $u$ receives interest \name{/A/B/Y} (I2), LPM fails, thus \pfs runs a DFS at node B and finds a forwarding face $f1$, at node D. If $f_1$ is not a {\em local} face then \pfs forwards the interest to $f_1$, otherwise it sends a {\tt NOROUTE NACK} to the consumer. Finally, if  $u$ receives interest \name{/A/H/Z} (I3), \pfs runs DFS at node H, which fails to find any leaf in the sub-trie, resulting in sending a {\tt NOROUTE NACK} to the consumer.


Note that \pfs uses DFS when longest prefix matching fails. While this functionality may introduce an extra delay overhead, we argue that the DFS search of a leaf is fast with a major benefit consisting of preventing unnecessary discovery flooding. The \pfs algorithm will create a FIB with that prefixes are implicitly aggregated (\npm).

\begin{algorithm}
	\caption{\PFS (\pfs) algorithm when receiving an interest with name prefix $p_i$}\label{NPM}
\begin{algorithmic}[1]
\REQUIRE $p_i$    



\STATE $Sn \gets LPM('/')$

\IF{$IsLeaf(Sn)$}   
    \STATE Return $Sn.f$ \COMMENT{LPM found a face}
\ELSIF{$f \gets DFS(Sn)$} 
    \IF{$isLocal(f)$} 
        \STATE Return ${\tt NOROUTE}$ \COMMENT{DFS successful but producer}
    \ELSE
        \STATE Return $f$ \COMMENT{DFS successful, forwarder}
    \ENDIF
\ELSE 
    \STATE Return ${\tt NOROUTE}$ \COMMENT{\pfs: No face found in sub-trie}
\ENDIF
        
\end{algorithmic} 
\end{algorithm}

\subsection{Multipath Discovery: Finding Alternate Routes} \label{MPALG}


As described in the previous section, \npm achieves an implicit aggregation of the FIB and maintains a smaller FIB size by trying approximate faces (\ie routes). However, when these faces fail to reach the correct producer (\ie either because the wrong producer is reached or a node with no face is encountered), a {\tt NOROUTE NACK} received by the consumer initiates an interest discovery to store a new, and more accurate, route at the FIB. 

\sol implements a multipath feature allowing any node to exhaust all forwarding faces prior to forwarding the {\tt NOROUTE NACK} back towards the consumer, thus reducing the overhead of initiating a new broadcast-based discovery. Note that storing multiple faces at the leaf nodes does not increase the size of the FIB, but simply adds few bytes to only leaf nodes.


\sol implements a NACK handling algorithm, as described in Alg.~\ref{nack-handling}, which checks the FIB for another alternative route for the same given prefix $p_i$ and if say a node $u$ successfully finds a face, $f$, for $p_i$, $u$ informs all nodes downstream and the consumer of the existence of an alternative route which was not tested. Consumer then can quickly (\ie without waiting for a timeout), re-transmits the same interest which will follow the same path until reaching node $u$, which can try the alternative route towards the producer. This process can continue until all nodes exhausts all available faces leading to producer or abruptly interrupted by consumer, after few attempts, to send a discovery interest and store a new route to producer. Alg.~\ref{nack-handling} allows a given node $u$ to change the NACK reason to {\tt ALT\_ROUTE} to inform all nodes downstream of the existence of an alternative, un-tested route.   

This multipath feature, reduces the need to broadcast many interest discovery messages which will be flooded, thus adding a major overhead and unnecessary congestion. We will evaluate the impact of this feature on latency and overhead in the evaluation section.


\begin{algorithm}
	\caption{OnReceiveNACK: \sol's multipath feature $<NACK_{p_i}>$ from face
 $oFace$}\label{nack-handling}
	\begin{algorithmic}[1]
 
\REQUIRE $NACK_{p_i}$, $oFace$     
\IF{ $Reason(Nack_{p_i})==$ {\tt NOROUTE} }
    \STATE $RemoveFaceFromLeaf(p_i, oFace)$ \COMMENT{Remove failed face and check if another one exists}
    \IF{$f \gets ExistAnotherFace(p_i)$} 
        \STATE $SetNackReason(NACK_{p_i},$ {\tt ALT\_ROUTE}$)$ \COMMENT{If another face exists, inform downstream}
        \STATE $Send(NACK_{p_i},iFace)$
        \label{nack.DISCOVERED}
    \ELSE
        \STATE $RemoveFIBEntry(p_i)$ 
        \label{nack.expire}
        \STATE $Forward(NACK_{p_i})$
    \ENDIF
\ELSIF {$Reason(Nack_{p_i})==$ {\tt ALT\_ROUTE}}
    \STATE $Forward(NACK_{p_i})$ \COMMENT{Nodes receiving {\tt ALT\_ROUTE} NACK keep their face to test alternative route}
\ENDIF

\end{algorithmic} 
\end{algorithm}



Now, we describe how \sol discovers and stores multiple routes for the same prefix. Note that discovery interests are broadcasted and do not follow LPM or \pfs mechanisms. We use a similar approach to Shi \etal's approach which implements a temporary data structure to save the PIT record for a given time allowing multiple Data messages are to be received for the same interest discovery~\cite{liang2020enabling}. While Shi \etal's approach works well for paths with two or more egress links, it fails when two paths have a node with two ingress links. This failure is due to the loop prevention mechanism which prevents any node of accepting two interests with the same nonce~\cite{afanasyev2014nfd}. \sol overwrites this rule, if and only if, it receives a discovery interest. \sol allows multiple interest discoveries to be received, their incoming faces added to PIT, however only the first interest is forwarded/broadcasted to the upstream faces. This change will not create looping interests (\ie interests forwarded indefinitely in the network), but yet allows a data received by the node to be sent to all incoming faces which help discover multiple routes instead of one. To allow that, when an interest discovery is received by a forwarder, and it is already in the PIT table\footnotetext{Same prefix and same nonce, which we refer to as interest (I) in Alg.~\ref{Id-handling}}, the forwarder node adds the interest incoming face (iFace) to the corresponding PIT record and silently discards the received duplicate interest (Alg.~\ref{Id-handling}).

\begin{algorithm}
	\caption{OnReceiveInterestDiscovery $<I>$ from $iFace$}\label{Id-handling}
	\begin{algorithmic}[1]
\REQUIRE $I$, iFace     
\IF{ $ExistsInPIT(I) ==0$ }
    \STATE $InsertPIT(I, iFace)$ \COMMENT{First discovery interest is saved and broadcasted to all faces except iFace}
    \STATE $BroadcastToAllFaces(I, oFaces)$ \COMMENT{Broadcast discovery to all outgoing faces, oFaces}
\ELSE
    \STATE $AddIncomingFace(I, iFace)$ \label{lg.aggdiscovery} \COMMENT{Append incoming face for duplicate discovery interests}
    \STATE $DiscardInterest(I)$ \COMMENT{Duplicate discovery interests are dropped}
\ENDIF
	\end{algorithmic} 
\end{algorithm}


On receiving a discovery Data message (\ie a Data with a discovery tag), a forwarder node, $u$, forwards the message to all downstream paths, iFaces, which create multiple ingress paths when $|iFaces|>1$. To help create multiple egress path, $u$ forwards the first Data message $D$ downstream and removes all incoming faces, iFaces, as well as the outgoing face where the Data is received from, oFace, and set an expiry time for the PIT record. Note that \sol keeps the PIT alive and all outgoing faces which the node did not receive any Data from yet, for potential alternative route announcement on those outgoing faces. This mechanism, similar to the one proposed by Shi \etal\cite{shi2017broadcast}, allows discovering multiple egress paths. Alg.~\ref{Id-handling} and Alg.~\ref{Dd-handling} provide details on how interest discoveries and Data discoveries are handled by \sol respectively.


\begin{algorithm}
	\caption{OnReceiveDiscoveryData $<D>$ from interface $oFace$}\label{Dd-handling}
	\begin{algorithmic}[1]
\REQUIRE $D$, $oFace$     

\IF{$iFaces \gets  PIThasFace(D)$} \label{alg.dataDall} 
    \STATE $FIBinsert(D,oFace)$\COMMENT{Data from first path arrived}
    \STATE $SendDataDownstream(D,iFaces)$ 
    \STATE $RemoveFacePIT(iFaces,oFace)$ \COMMENT{Keep outgoing faces which did not send data yet}
    \STATE $PitRecordExipreTimer(D,tmp)$ \COMMENT{Set a timer $tmp$ for the PIT record to expire} 
\ELSIF{$tmp >0 $}  
    \STATE $FIBinsert(D, oFace)$ \COMMENT{Insert an alternative egress path}\label{alg.addmore}
    \STATE $RemoveFacePIT(D,oFace)$
\ELSE 
    \STATE $sendNACK(${\tt UNSOLICITED\_DATA}$)$
\ENDIF

	\end{algorithmic} 
\end{algorithm}

\subsection{\sol's Consumer Stop-and-Wait Mechanism}\label{stop-wait}

Most applications send multiple messages to the same producer. To avoid sending numerous interest discovery requests as soon as an application sends an interest to a given producer, \sol implements a Stop-and-Wait consumer mechanism.  This mechanism queues interests for any given prefix that has an ongoing discovery. Using this mechanism, consumers do not send multiple interests discoveries for the same prefix name, thus avoiding unnecessary overhead caused by sending multiple ``redundant flooding messages'' to discover the same route.  

\begin{figure}
     \centering
         \centering
         \includegraphics[width=0.65\linewidth]{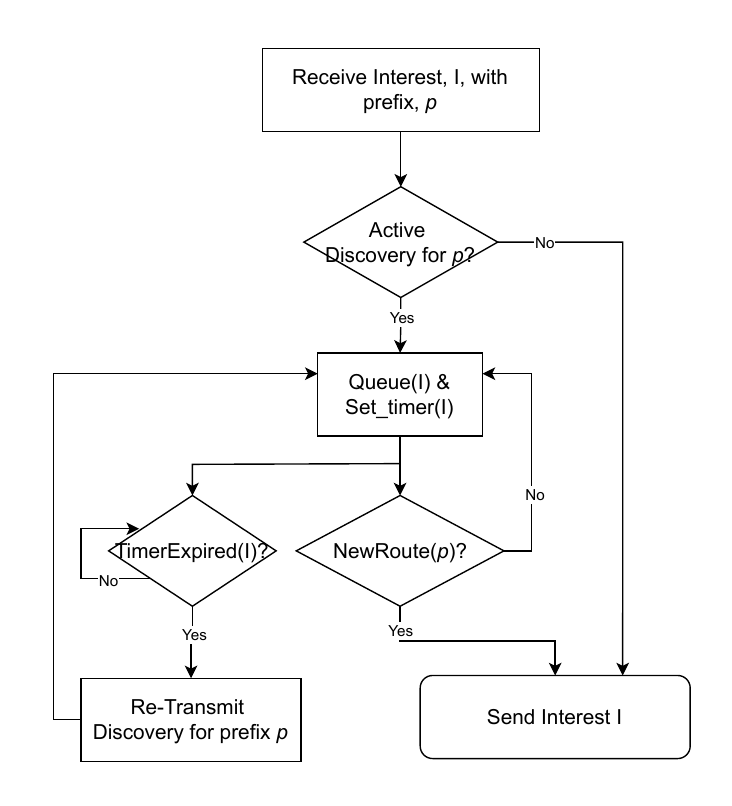}
     \caption{Flowchart diagram of \sol's Stop-and-Wait mechanism. Interests are queued waiting for the creation of a new route or a timeout of a  current active discovery. }
        \label{fig:stopNwait}
 \end{figure}

\sol's Stop-and-Wait mechanism, as depicted in the flowchart in Figure~\ref{fig:stopNwait}, aims at queue all interests sharing the same prefix $p$ with an ongoing current discovery. The queued interests wait until: (1) a new route is discovered and stored in the FIB, or (2) a timer expiration triggering a new discovery process for the same prefix $p$.   

\sol's consumer, initiates a route discovery for prefix $p$ as soon as it receives {\tt NOROUTE} NACK, and starts a timer for the requesting interest with prefix $p$. Any new interest received by the consumer app sharing the same interest $p$, will follow the Stop-and-Wait mechanism described above (Figure~\ref{fig:stopNwait}). When a new route is discovered, all queued interests are sent with the highest priority (\ie prior to new interests). 




\section{Evaluation}\label{seceval}

In this section, we compare the performance of \sol against state-of-the-art \SL algorithm~\cite{liu2017unified}. First, we describe our evaluation methodology, metrics, and then we study the impact of various parameters. on \sol's performance.

\subsection{Simulation Setup}
We use ndnSIM~\cite{afanasyev2012ndnsim}, a module of ns-3~\cite{henderson2008network}, to implement and evaluate \sol. We perform our simulation on a Desktop machine with a 4 core-i7 Intel CPU and 8GB memory.

\noindent{\bf Network Topology:} We created an ISP-like network topology consisting of $N$ nodes, including $R$ routers, $C$ Prefixes (consumers), and $P$ Producer Connection Points. The $R$ routers consists of $R_c$ core and $R_e$ edge routers. While the core routers are connected to three other core routers, edge routers however are connected randomly to 1, 2, or 3 core routers (and to none of the edge routers). Edge routers are also used as access nodes for consumers and producers. We connect randomly each consumer and producer to one and only one edge router. While we vary $C$ and $P$ in our simulation, we choose to fix $R_e=16$ and $R_c=21$. In our topology the average path length between any consumer and producer is roughly 3.2 hops. 

\noindent{\bf Implementation:} We implement  \pfs  forwarding strategy at each node in the network, including forwarders, producers, and consumers. Consumers also implement \sol's Stop-and-Wait mechanism as described in section~\ref{stop-wait}. Consumers' apps, starting at a random time spanning from beginning of simulation and 50 seconds after, send requests at a rate of 8 interest per second. Each consumer $i$ sends interests with the same prefix $p_i$, for a given producer $u$. interests have the following name format: \name{/$p_i$/seq}, where seq is the sequence number for consumer $i$. We set all producers to produce a total of $M \ge P$ prefixes, such that each producer can produce one or many prefixes. In our simulation, we disable in-network caching to evaluate the features introduced by \sol without any unknown parameter such as caching.

We perform twenty simulation runs for every experiments where we vary the topology and the producer prefix association. Each simulation run is set to 60 seconds.

\subsection{Evaluation Metrics}\label{metrics}
We use the following three metrics to evaluate \sol:


\begin{enumerate}
    \item \emph{Average number of FIB entries}: We measure the number of FIB entries for a given router as the number of leaves in its FIB trie. This metric is proportional to the number of nodes in the trie. 
    We measure the average number of FIB entries for all routers as well as for core-only. 
    \item \emph{Network overhead}: Measured as the number of broadcasted discovery interests (\ie interests with a discovery tag).  
    \item \emph{Average number of redundant paths}:  We measure the number of disjoints paths for each prefix at each router and compute the average per prefix and per router. 
    
\end{enumerate}

\subsection{Results and Analysis}

We design \sol to reduce the FIB size and thus reduce the lookup latency as shown in Figure~\ref{lookuptime}. We first compare the FIB sizes of \sol and \SL. 
 
\subsection{\sol's FIB Table Size} \label{exp-ISP}
We vary the number of prefixes per producer to evaluate how these two algorithms, \sol and \SL, construct their FIB tries and which one scales better.

\begin{figure}
    \centering
    \subfloat[Avg. FIB records in core+edge routers]{\includegraphics[width=0.47\linewidth]{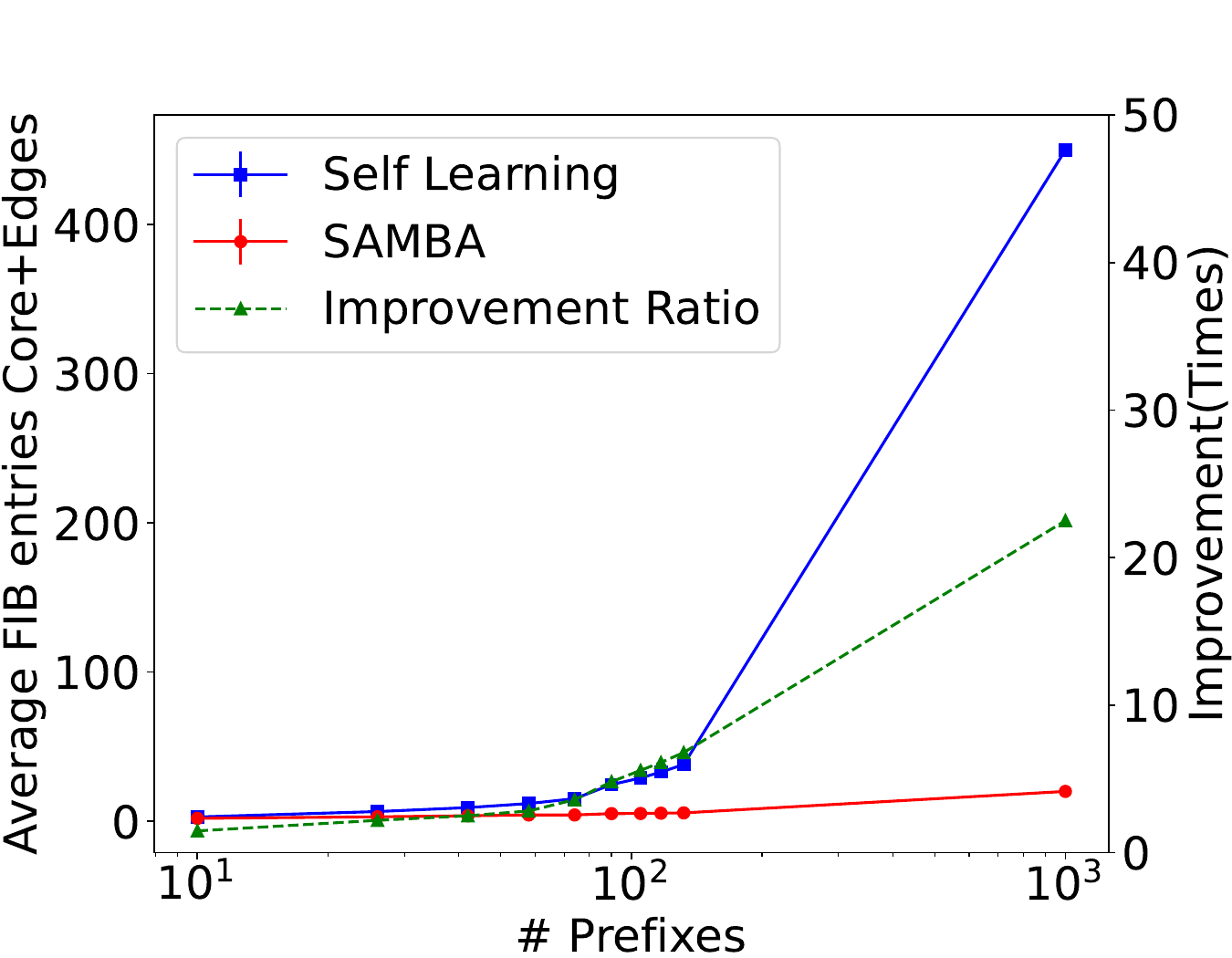}
    \label{SvM-FIB-core-edge}}
    \hspace{0.01\linewidth} 
    \subfloat[Avg. FIB records in core routers]{\includegraphics[width=0.47\linewidth]{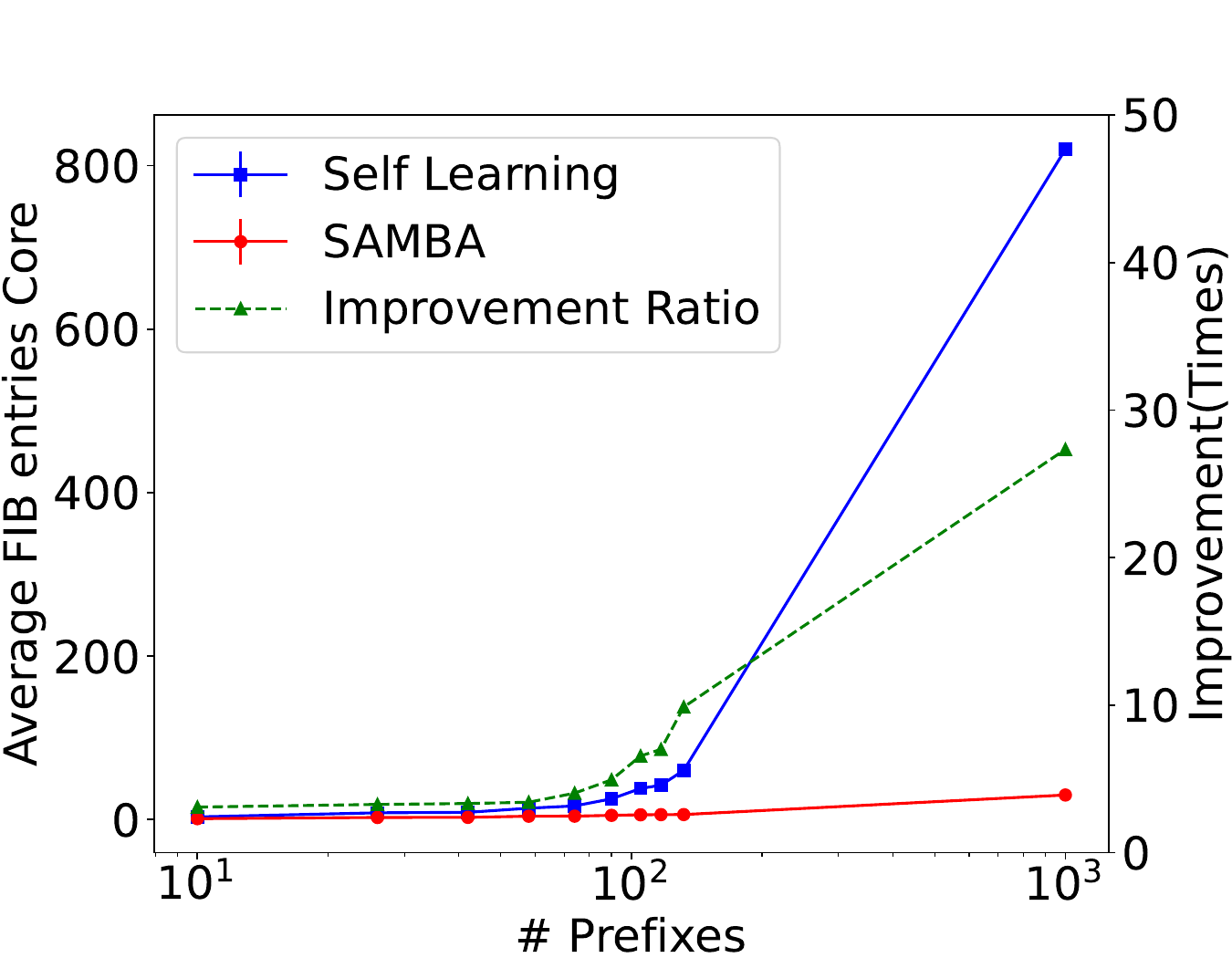}
    \label{SvM-FIB-core}}
    \caption{Comparison of \sol's and \SL's average FIB table size as the  number of prefixes, $C$, increases. The improvement ratio, in the right y-axis, increases as the number of prefixes per producer increases (x axis in logscale).}
    \label{fig:SizeVSm}
\end{figure}

We first, fix the number of producers in the network, $P=4$, and vary the number of prefixes, $C$. As $C$ increases, producers serve more and more prefixes (\eg a google server can serve Gmail, Drive, Calendar, and Photo services). We plot, in Figure~\ref{fig:SizeVSm}, the average FIB size of all nodes in the network (Fig.~\ref{SvM-FIB-core-edge}) and for core nodes only (Fig.~\ref{SvM-FIB-core}), using \sol and \SL, as we increase the number of requested prefixes, $C$. 
We show that \sol  outperforms \SL and keeps the size of the FIB smaller as $C$ increases. \sol's FIB size does not exceed 22 and 35 prefixes on average for all routers and core routers respectively, while \SL's FIB size increases exponentially exceeding 420 and 805 prefixes respectively.

Additionally, we measure the improvement ratio measured as the fraction of average number of FIB records in \SL to average number of FIB records in \sol, and show that \sol's improvement increases as the number of prefixes per producer increases. In fact, \sol re-uses the same saved route for the producer, thus the more prefixes served by unique producer the better performance achieved by \sol. \SL, however, re-discovers all paths to all prefixes and thus the size of its FIB trie (\ie or table) increases exponentially. The improvement ratio increases from 2$\times$ and 4$\times$ when $C=10$, to 20$\times$ and 30$\times$ when $C=1k$ for all routers and core routers respectively. This improvement gain increases shows how \sol can scale better compared to \SL, thus reduces the trie search and lookup times as shown in Figure~\ref{lookuptime}.



\begin{figure}[tbp]
    \centering 
        \subfloat[Avg \# FIB records in core+edge when $P$ changes]{\includegraphics[width=0.47\linewidth]{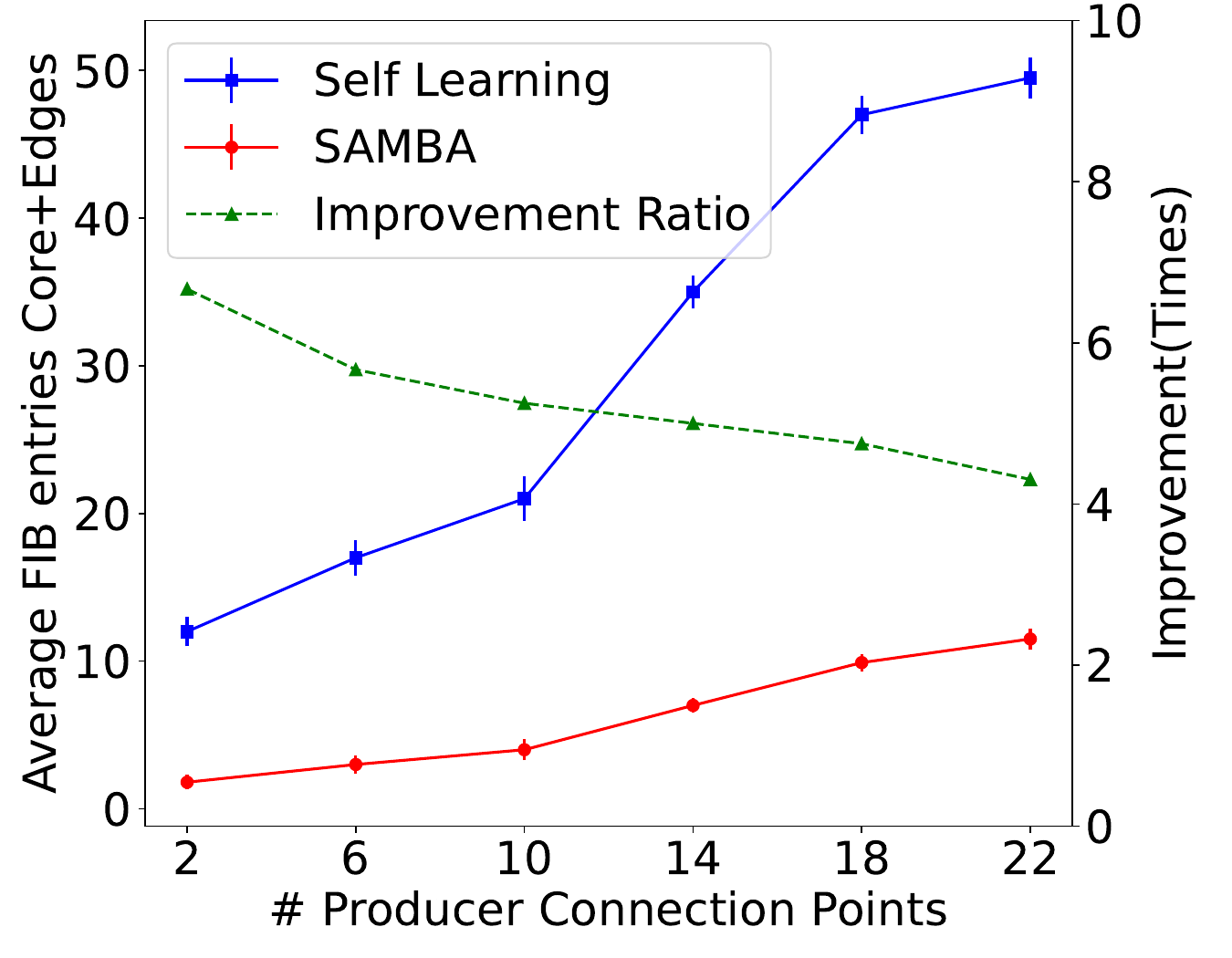}
        \label{exp2-FIB-core-edge}}
        \subfloat[Avg \# FIB records in core routers when $P$ changes]{\includegraphics[width=0.47\linewidth]{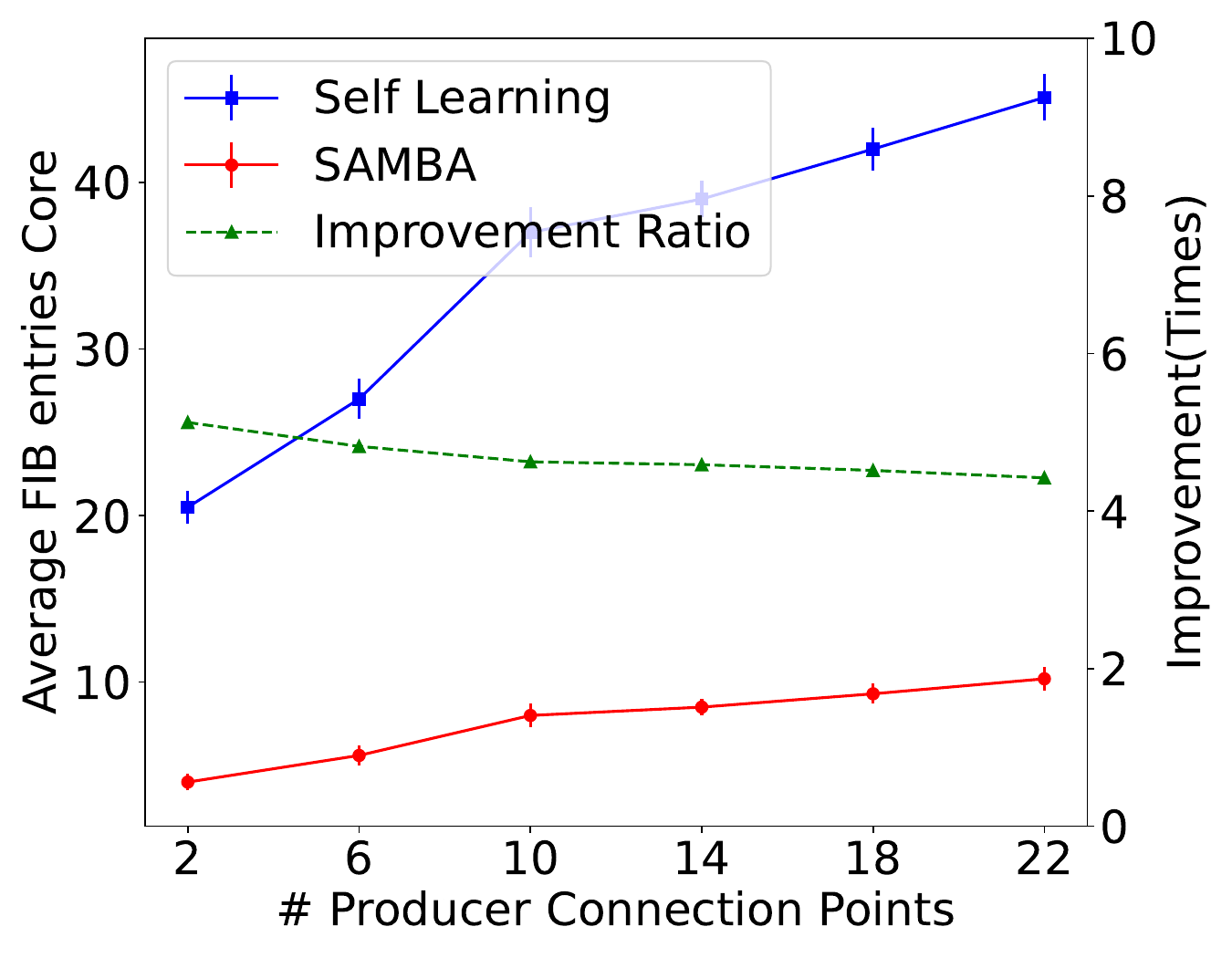}
        \label{exp2-FIB-core}}
    \caption{Comparison of \sol's and \SL's average FIB table size as the number of number of producer connection points), $P$, increases. The improvement ratio increases  as the number of prefixes per producer increases. }
    \label{fig:sizeVSp}
\end{figure}


We also fixed the number of prefixes, $C=100$, and vary the number of producers connection points, $P$, to investigate how \sol scales as the number of producers increases. Our results in Figure~\ref{fig:sizeVSp} show that while \sol keeps the FIB size small compared to \SL, the improvement ratio decreases as $P$ increases. The improvement ratio for all routers decreases from 6$\times$ to 6$\times$ as $P$ increases from 2 to 22. However, this improvement decreases much less for core routers (from 5$\times$ to 4.3$\times$), which are more important in any networking architecture, thus keeping their FIB smaller is more critical than edge routers. 
As the number of prefixes per producer connection points decreases--\ie $P$ increases, \sol's \pfs algorithm will fail to use accurate approximate routes resulting in more and more discoveries. However, we note that these discoveries remain smaller than those initiated by \SL. 

\subsection{Overhead Analysis} \label{exp-overhead}

In addition to \sol's main objective to reduce the FIB size, we argue that \sol also uses much less message overhead when compared to other state-of-the-art routing schemes. While proactive solutions have much higher overhead, we compare \sol to the \SL reactive approach to quantify when and how \sol utilizes fewer discoveries to achieve the same delivery performance.

\begin{figure*}
    \centering
    \subfloat[Interest Discovery  when $C$ changes]{\includegraphics[width=0.25\linewidth]{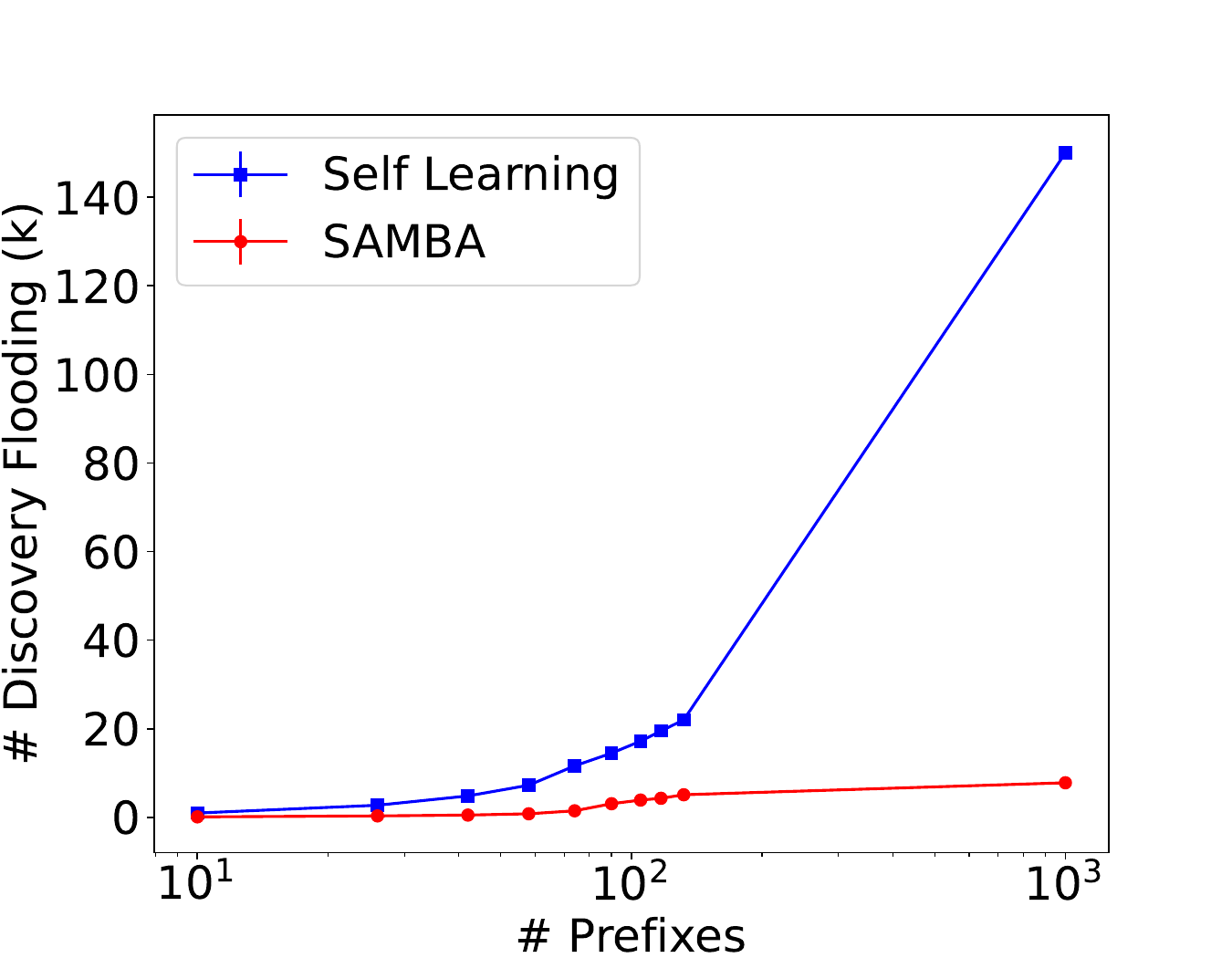}
    \label{exp1-network-discovery}}   
    \subfloat[Interest Discovery sent to the Network when $P$ changes]{\includegraphics[width=0.25\linewidth]{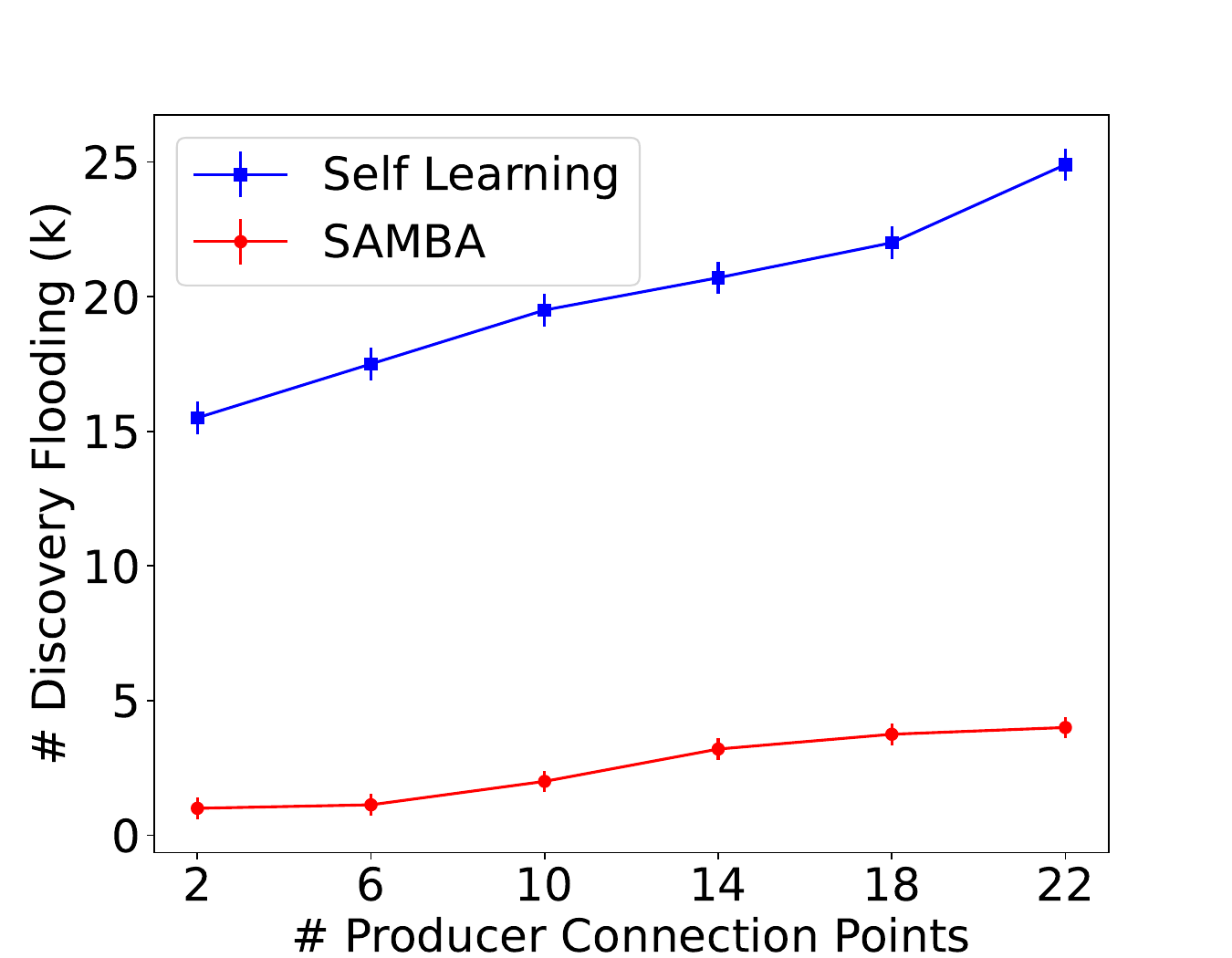}
        \label{exp2-network-discovery}}
    \subfloat[Network overhead of  multipaths when $C$ changes]{\includegraphics[width=0.25\linewidth]{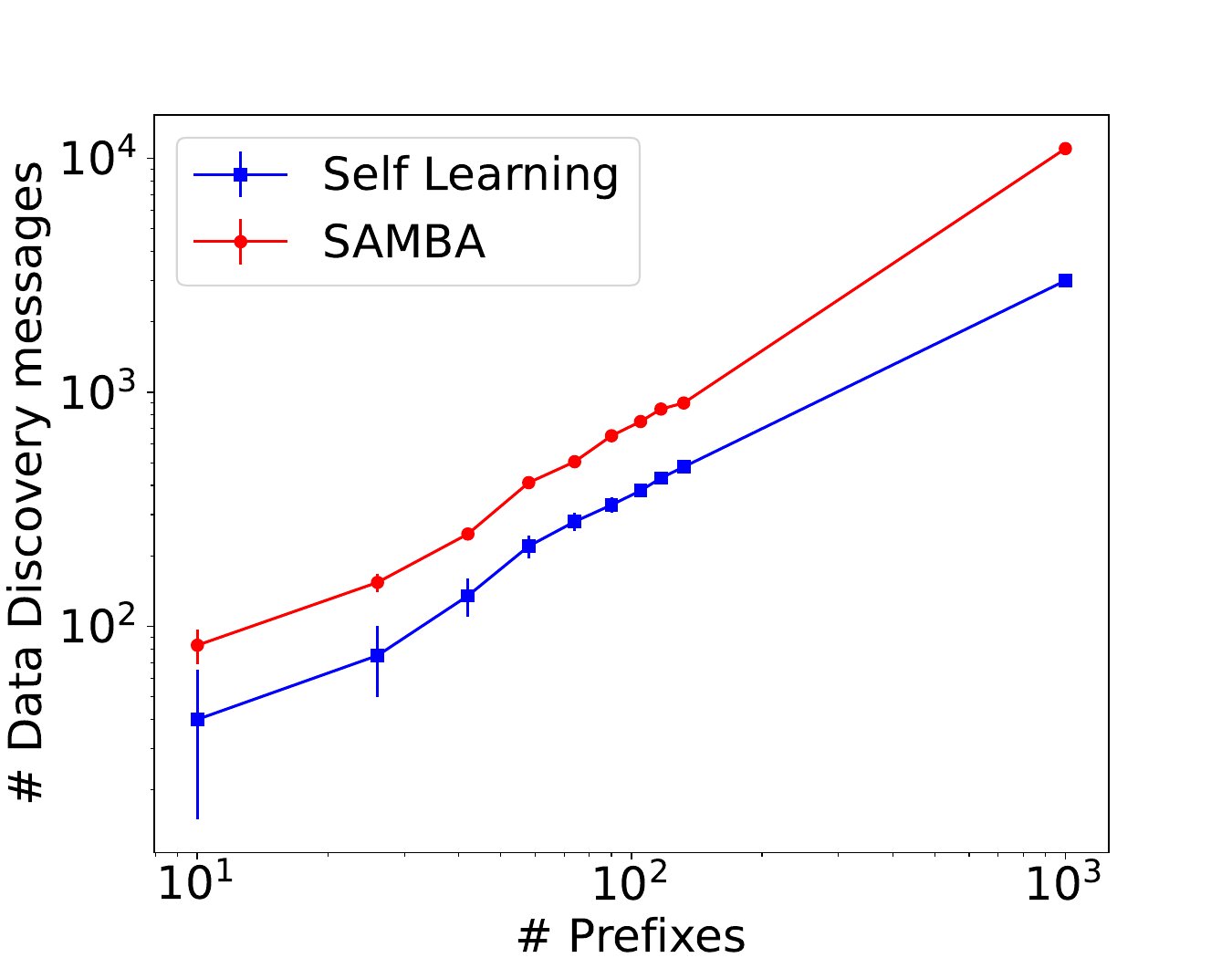}
    \label{exp.data_discovery}}   
    \caption{Network overhead comparison of \sol and \SL.}
    \label{fig:exp1}
\end{figure*}

However, we measure the networks overhead as number of interest discovery messages that flooded over the network, in both experiments that is shown in section \ref{exp-ISP}, \sol has lower overhead  compared to \SL. Figure \ref{exp1-network-discovery} shows that \sol is able to reuse FIB records around 20 times more than \SL. It, therefore, generates less overhead than \SL, especially when the number of requested different prefixes, $C$, increases. For instance when $C=1k$, \SL flooded 150k interests to discover new routes, however, \sol did not send more than 8,087 interests and reused most of the pre-existing FIB records to forward consumers' interests.  After reusing pre-existing FIB records, \sol failed in only 10\% interest and ended up sending new discovery interests to add new routes.

Also, number of discovery messages in the second experiment when number of producer connection points $P$ is varying, increases for both \sol, and \SL, however, \sol has better results in decreasing the number of discovery flooding in the network as it is depicted in Figure~\ref{exp2-network-discovery}.

Other overheads of \sol is the delay and additional discovery interest due to wrong forwarding. While the wrong forwarding happens just in the first interest of a connection and the delay of finding a new path is less than a RTT, so these overheads is not considerable and it is still fewer than number discovery messages in \SL.

We also evaluate the overhead of \sol for detecting the multipaths  by measuring number of data discovery messages that return on the router link. While in \SL just a path is discovered, and duplicate discovery interests from multiple paths dropped, \sol discovers all available paths, consequently number of data discovery message increases (Figure~\ref{exp.data_discovery}). This additional overhead of \sol is doubled compared to \SL. However, we will show how alternative paths can increase the \sol performance in presence of link failure.

\subsection{\sol's Performance in Presence of Link Failures} \label{exp-sharedpath}

In this section, first, we compare level of path  redundancy in \sol and \SL. To this end, we measure the Average number of Path per Prefix (APP) in the core and edge routers respectively. In this experiment we use the network topology of section \ref{exp-ISP}, fixed the number of prefixes $C=30$, and producers $P=4$. Then, we vary number of core and edge routers link $k$ from one to 10.  

\begin{figure}
    \centering
    
    \subfloat[Avg \# discovered paths for 30 consumers in core routers]{\includegraphics[width=0.45\linewidth]{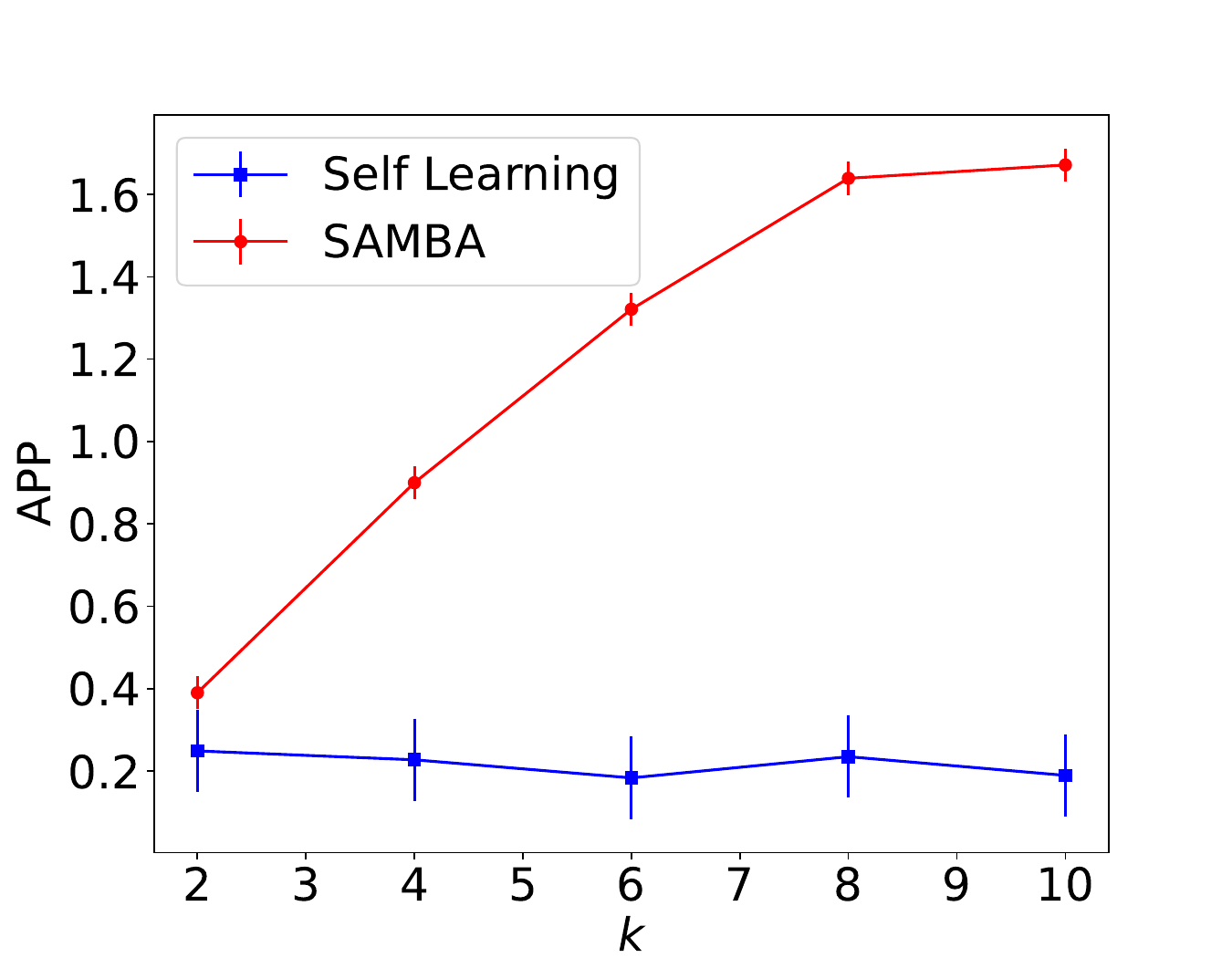}
        \label{exp3-FIB-core}}
    \hfill
    \subfloat[Avg \# discovered paths for 30 name consumers in core and edge routers]{\includegraphics[width=0.45\linewidth]{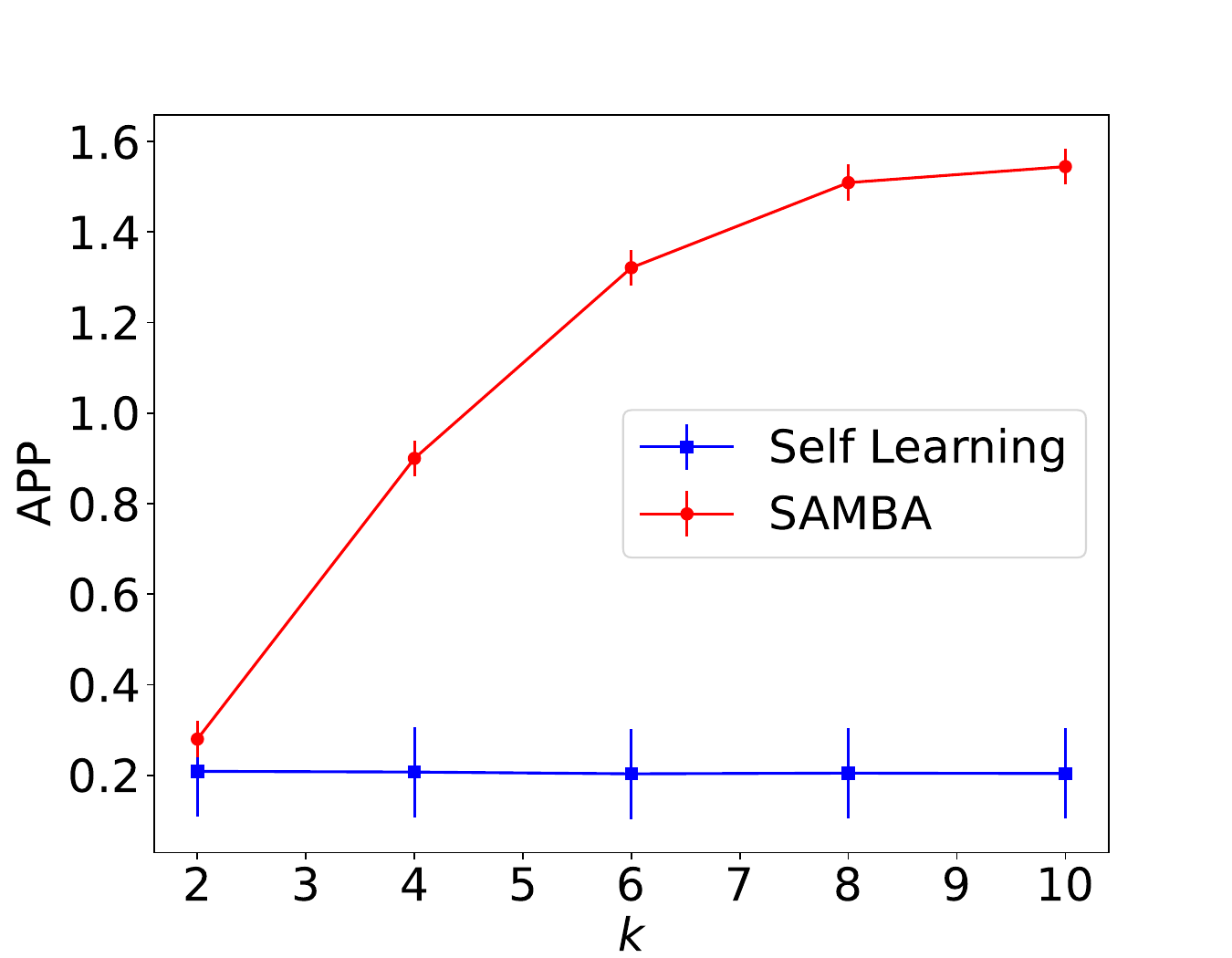}
        \label{exp3-FIB-core-edge}}
    
    \caption{In this experiment, the level of redundancy increases by adding a degree of $k$ in core and edge routers. While detected paths in \SL are almost the same, in \sol, it increases by elevating redundant links in core and edge routers.}
    \label{fig:exp3}
\end{figure}

While \SL can detect only one path per discovery and each router just has a next hop per prefix, \sol can discover all existing paths. With maximum of $k=10$ for each router \sol can discover 1.67 APP in core routers (Figure~\ref{exp3-FIB-core}), while \SL can just discover on average almost 0.25 APP for all $k$ values, also \SL can discover 1.53 APP in core and edge routers (Figure~\ref{exp3-FIB-core-edge}), and \SL can discover 0.2 APP.

 \begin{figure}
    \centering
    
    \subfloat[Consumer C0 requests /P1.]{\includegraphics[width=0.45\linewidth]{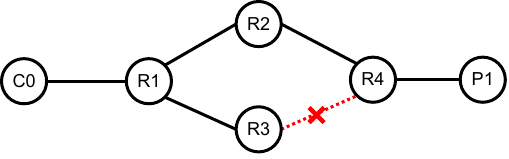}
        \label{fig:exp4}}
    \hfill
    \subfloat[Consumer throughput when, AIMD used ]{\includegraphics[width=0.45\linewidth]{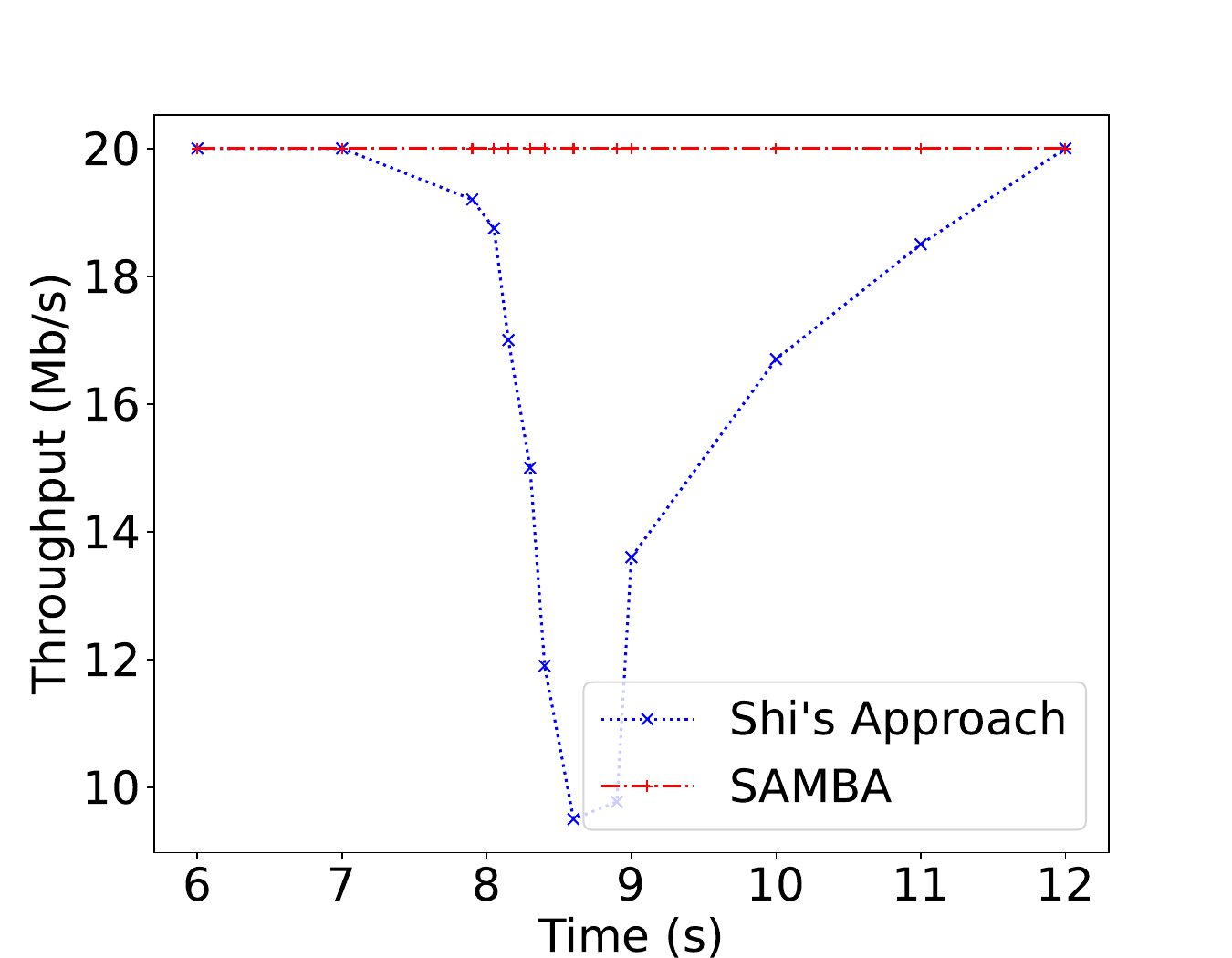}
        \label{fig:exp4-result}}
    
    \caption{Link Failure scenario when R3-4 Link fails in second 8, \sol uses the alternate route immediately.}
    \label{fig:exp4-all}
\end{figure}

\noindent{\bf Toy Scenario with Link Failure}
In this experiment We simulate link failure scenario with a simple topology shown in Figure \ref{fig:exp4} with two disjoint paths towards producer P1, and each link delay is 10ms. We run the simulation for 12 seconds and simulate an R3-R4 link failure at time 8 seconds using the $LinkDown$ function in ndnSIM.  In this experiment, we implement consumer C1 requests prefix \name{/P1} from P1 and runs congestion control mechanism using AIMD~\cite{schneider2016practical} algorithm, also we set starting window size to one.To find failed links, we exploit BFD \cite{katz2019bidirectional} protocol which is a periodic bidirectional way in link layer, with time interval $Int=5 ms$, and dead interval of 3.


Figure~\ref{fig:exp4-result} shows the throughput measured at C1 node over time. We compare \sol with Shi's approach \cite{liang2020enabling} that can discover multipath through different producers, while in Shi's approach consumer throughput decreases by up to 50\% when the link fails. It took the consumer more than 1.5 seconds to re-discover a new path and recover its maximum throughput. Consumer's congestion window got decreased and the delay to recover is a result of a slow start/congestion avoidance mechanism. Whereas, \sol's consumer does not get impacted by the link failure and maintains a maximum throughout during  simulation. \sol uses its multipath feature to set two disjoint paths (when available, \eg at R1) resulting in a seamless switching to the second path C0-R1-R2-R4-P1 as soon as R1 receives a {\tt NOROUTE NACK} to divert traffic towards R2 instead of R3, and R1 sends {\tt ALT\_ROUTE} to C1, so the consumer will not activate the slow start state. Note that in case the second path fails, \sol will then initiate a route discovery and set new paths towards P1 (this scenario is not simulated in our experiment).
In addition to the throughput gains, \sol will also reduce the number of discovery messages sent because of alternative paths while Shi's approach has exactly 195 discovery messages to recover from the link failures, which \sol avoided.

         

\label{secconslusion}



\section{Conclusion and Future Work} \label{conclusion}
In NDN, many routing approaches face scalability issues due to the increasing number of FIB records affecting lookup times. To address this, we developed \sol, which \pfs and \npm to maintain small FIB tables by finding the nearest prefix for a given name. \sol also features multipath discovery for route redundancy and a Stop-and-Wait mechanism to minimize redundant discoveries. Evaluations show SAMBA achieves 20 times fewer FIB records and 5 times fewer flooding messages compared to Self Learning. Future work will focus on implementing \sol on a real testbed and enhancing its multipath management.

\bibliographystyle{ieeetr}
\bibliography{biblography}
\end{document}